\documentclass[fleqn,usenatbib]{mnras}

\usepackage[T1]{fontenc}
\usepackage{ae,aecompl}

\usepackage{graphicx}	% Including figure files
\usepackage{amsmath}	% Advanced maths commands
\usepackage{amssymb}	% Extra maths symbols
\usepackage{siunitx}
\usepackage{color} % gestion de différentes couleurs

\newcommand{\pderiv}[2]{\frac{\partial #1}{\partial #2}} % write the partial derivative as d#1/d#2
\newcommand{\dif}[1]{\mathrm{d}#1\,} % write the differential d#1

\newcommand{\vnabla}{\vec{\nabla}}
\newcommand{\vnablapara}{\vec{\nabla}_\parallel}
\newcommand{\vj}{\vec{j}}

\newcommand{\vb}{\vec{b}}
\newcommand{\vi}{\vec{i}}
\newcommand{\vJpara}{{\vec{J}_\parallel}}
\newcommand{\Tdh}{T_{\mathrm{dh}}}
    
\newcommand{\vuphi}{\vec{u}_\phi}
\newcommand{\vuthet}{\vec{u}_\theta}
\renewcommand{\vec}[1]{\mathbf{#1}}
\title[Spider heat redistribution]{A model for redistributing heat over the surface of irradiated spider companions}

\author[G. Voisin et al.]{
	Guillaume Voisin$^{1,2}$\thanks{E-mail: guillaume.voisin@manchester.ac.uk;\newline astro.guillaume.voisin@gmail.com},
	M. R. Kennedy$^{1}$,
	R. P. Breton$^{1}$,
	C. J. Clark$^{1}$,
	D. Mata-S\'anchez$^{1}$
	\\
	$^{1}$ Jodrell Bank Centre for Astrophysics, Department of Physics and Astronomy, The University of Manchester, Manchester M19 9PL, UK\\
	$^{2}$ LUTH, Observatoire de Paris, PSL Research University, 5 Place Jules Janssen, 92195 Meudon, France
}
\date{Accepted XXX. Received YYY; in original form ZZZ}

\pubyear{2020}

\begin{document}
\label{firstpage}
\pagerange{\pageref{firstpage}--\pageref{lastpage}}
\maketitle

\begin{abstract}
Spider pulsars are binary systems containing an energetic millisecond pulsar that intensely irradiates a closely orbiting low-mass companion. Modelling their companion's optical light curves is essential to the study of the orbital properties of the binary, including the determination of the pulsar mass, characterising the pulsar wind and the star itself.
We aim to generalise the traditional direct heating model of irradiation, whereby energy deposited by the pulsar wind into the stellar envelope is locally re-emitted, by introducing heat redistribution via diffusion and convection within the outer stellar envelope.
We approximate the irradiated stellar envelope as a two-dimensional shell. This allows us to propose an effective equation of energy conservation that can be solved at a reduced computational cost. We then implement this model in the \texttt{Icarus} software and use evidence sampling to determine the most likely convection and diffusion laws for the light curve of the redback companion of PSR J2215+5135.
Redistribution effects concentrate near the terminator line of pulsar irradiation, and can create apparent hot and cold spots. Among the models tested for PSR J2215+5135, we find that all models with heat redistribution are more likely than symmetric direct heating. The best-fitting redistribution model involves diffusion together with a uniformly rotating envelope. However, we caution that all models still present serious systematic effects, and that prior knowledge from pulsar timing, spectroscopy and distance are key to determine with certainty the most accurate redistribution law.
We propose an extension of the direct heating framework that allows for exploring a variety of heat redistribution effects. Future work is necessary to determine the relevant laws from first principles and empirically using complementary observations.
\end{abstract}


%%%%%%%%%%%%%%%%%%%%%%%%%%%%%%%%%%%%%%%%%%%%%%%%%%

%%%%%%%%%%%%%%%%% BODY OF PAPER %%%%%%%%%%%%%%%%%%
%-------------------------------------------------------------------
\section{Introduction}
% Problem : modleling of spider pulsar and mass controversy 
% wealth of informtuion 
% models: spots, b ducting, intrabinary shock, conv
% icarus
% asymmetry J2051
% in this paper we present a generalisation by assuyming and application 

Spider pulsars are binary systems in which the primary component is a millisecond pulsar and the secondary a low-mass star, which we will call the companion in this paper. The orbital period of the binary is typically of a few hours. The companion is generally close to filling its Roche lobe and usually assumed to be tidally locked onto the neutron star. Spiders are found in two sub-species: redbacks with companion mass $\gtrsim 0.1 M_\odot$, and black widows with a companion mass of a few $0.01 M_\odot$. These names were coined after two arachnid species which share the characteristic that the light male companion is sometimes eaten by the heavier female. For their stellar counterparts, there is indeed suspicion that the low-mass companion is being gradually evaporated by the intense wind of high-energy particles radiated by the pulsar \citep[e.g.][]{fruchter_millisecond_1988}. This is evidenced by the wide radio eclipses attributed to clumps of ablated material surrounding the companion far out of its Roche lobe, although it is as yet unclear whether this is sufficient to lead to the disappearance of the star   \citep[e.g.][]{polzin_study_2020}. 

This irradiation of the companion, which often exceeds the intrinsic luminosity of the star, results in a characteristic day-night pattern in its light curve as it moves around its orbit with the pulsar. Once modelled and combined with pulsar timing and potentially with spectroscopy, optical light curves allow one to infer the inclination and the mass ratio of a system, and thereby the mass of the two components. In particular, there is evidence that the mass of spider pulsars could be on average larger than for other pulsars \citep{linares_super-massive_2019, strader_optical_2019}, and so they could be used to constrain the maximum mass and equation of state of neutron stars \citep[e.g.][]{ozel_masses_2016}.
Modelling of optical observations of spider companions also provides an indirect probe of the pulsar wind. In particular, comparing the temperatures of the day and night sides of the star provides an estimate of the amount of irradiating power necessary to sustain such a difference. Modelling the interaction of the wind with the stellar material, and in particular determining what components (gamma rays, leptons, hadrons) can penetrate below the photosphere and produce the observable effective temperature difference also provides insight in the composition of the wind \citep{zilles_diagnosing_2019}. 

Both the determinations of the orbital and wind parameters are highly dependent on the modelling of the temperature at the surface of the companion star. A common approach consists in assuming a direct heating of the surface whereby the energy deposited by the pulsar wind is re-radiated by the companion at the exact location where it is absorbed \citep[e.g.][]{breton_koi_2012}. Although this approach permits reasonable fits of some light curves \citep[e.g.][]{breton_discovery_2013}, this model is unable to account for asymmetries between the leading and trailing edge of the companion (assuming a symmetric irradiation pattern) such as seen in, for example, the black widows PSR J2051-0827 \citep{stappers_intrinsic_2001} and PSR B1957+20 \citep{kandel_atmospheric_2020} or the redback PSR J2215+5135 \citep{romaniIBS, linares_peering_2018, schroeder_observations_2014}. 

Various models have been proposed to explain asymmetries. Two of them, the magnetic-field ducting of the energetic charged particles of the pulsar wind by the magnetic field of the companion \citep{romaniBDuct} and radiation from an intra-binary shock formed between the winds of the two components \citep{romaniIBS}, still assume direct heating of the companion but change the irradiation pattern from an isotropic point source through the interaction of the pulsar wind with the companion's own wind and/or magnetic field. A third approach consists in empirically adding hot or cold spots at the surface of the star \citep[e.g.][]{shahbaz_properties_2017}. Thus, in all cases heat is assumed to be neither diffused nor convected within the star after energy reaches the surface. % Could cite in prep J1227 + J1023 paper. 

In fact, the more general problem of irradiation of stellar atmospheres by a companion has been known and studied since the early twentieth century and is often referred to as the reflection effect (see \cite{claret_irradiated_2004} for a historical summary of the treatment of the problem). Approximated, perturbative treatments of irradiation-driven circulation in stellar atmospheres were derived by, in particular,  \citet{kirbiyik_circulation_1976},\citet{kippenhahn_hz_1979}, and \citet{kirbiyik_penetration_1982}. However these authors considered the case of irradiation caused by X-rays in main sequence stars, and not by gamma rays or high-energy particles, which are absorbed deeper in the atmosphere. X-ray irradiation on the other hand may not reach below the photosphere \citep[see, e.g.,][]{zilles_diagnosing_2019}. While it can still produce important changes in ionisation, as was shown observationally using phase-resolved spectroscopy \citep[e.g.][]{davey_irradiation_1992,phillips_outburst_1999,shahbaz_irradiation_2000}, these models are unlikely to suffice to explain as dramatic photometric changes as those seen in spider binaries due to the nature of the irradiating particles considered. For these reasons, their results cannot be immediately transposed to spider companions.

The aforementioned theoretical studies concluded that irradiation results in an atmospheric circulation that transports energy over the terminator of the irradiation pattern. It is to be noted that the amount of energy leaking to the dark side of the star remains limited since these models only consider perturbative irradiation, unlike what is observed in spider binaries. Besides, another perturbative study \citep{tassoul_meridional_1982} showed that the magnitude of the currents found in \citet{kirbiyik_circulation_1976},\citet{kippenhahn_hz_1979}, and \citet{kirbiyik_penetration_1982}, which are supersonic, might be largely overestimated due to the negligence of eddy viscosity in these models.

Numerical studies have also assessed the onset of circulation in irradiated binaries, starting with \citet{martin_application_1995}, using smoothed particles hydrodynamics \citep{eldridge_perturbations_2003}, or using full 3D hydrodynamics in order to capture the effects of the Roche-Lobe filling and Coriolis force \citep{beer_general_2002,beer_irradiation_2002}. These studies all concluded that a global circulation current was driven by irradiation, albeit with different properties. In particular, \citet{martin_application_1995} found a supersonic flow, similarly to the analytic work of \citet{kippenhahn_hz_1979} and \citet{kirbiyik_penetration_1982}, while \citet{beer_general_2002} found a subsonic flow. However, in the latter case, the slower flow velocity was not due to eddy viscosity as in \citet[][]{tassoul_meridional_1982} but to the possibility for to the stellar surface to deform. We also note that the topic of irradiated atmospheres has regained interest recently due to the discovery of the so-called hot Jupiters \citep[see, e.g.,][]{showman_equatorial_2011,showman_atmospheric_2018}.

In Sect. \ref{sec:model} we develop a simpler heat redistribution model within the outer layers of the star which constitutes a natural extension of direct heating models. In principle, the heat flux must be calculated from detailed stellar and atmospheric models of the star such as those summarised above. However, we focus in the present work on demonstrating the basic properties and the interest of this new framework by using simple diffusion-like and convection-like laws. To the best of our knowledge, diffusion-like laws are a novel feature of this type of model. In Sect. \ref{sec:application}, we apply this simple model to the light curve of the companion of PSR J2215+5135 \citep{linares_peering_2018} in order to empirically determine the most probable law. We then discuss the physical interpretation of the results in Sect. \ref{sec:disc}. At the time of finishing this paper, a similar model was published in \citep{kandel_atmospheric_2020} which appears to be a special case of the framework presented here, where no diffusion effect is considered and a particular convection law is used. For comparison purposes, we also reproduce this model in the present work. 

% compatible with all the other stuff
% The internal structure and evolution of the stellar companions is also unclear \citep{benvenuto, chen}. This evolution might highly depend on how the irradiation affects the surface of the star, for example 

\section{Superficial heat transport model }
\label{sec:model}
\subsection{Preliminary considerations: direct heating by high-energy particles}\label{sec:dh}
Currently, state-of-the-art light-curve-modelling softwares such as \texttt{Icarus} \citep{breton_koi_2012}, \texttt{ELC} \citep{orosz_use_2000}, \texttt{XRBCURVE}\citep{shahbaz_optical_2003}, or \texttt{BINSYN} \citep{linnell_binsyn_2012}, rely on the approximation that the power impinging on the companion star is thermalised  and re-radiated at the location on the photosphere where it was absorbed. This leads to the following energy balance,
\begin{equation}
\label{eq:directheating}
	\sigma \Tdh^4 = \sigma_{\rm sb} T_{\rm b}^4 +  L_{\rm w},
\end{equation}
where $\sigma_{\rm sb}$ is the Stefan-Boltzmann constant, $\Tdh$ is the temperature of the photosphere after irradiation, $T_{\rm b}$ is the base temperature without irradiation, $L_{\rm w}$ is the energy flux of the pulsar wind at the photosphere. 

Let us note that the base temperature $T_{\rm b}$ is not necessarily constant over the star, but can be affected by, for example, gravitational darkening or magnetic activity (star spots).
The irradiation flux $L_{\rm w}$ includes the cross-section of the stellar surface relative to the incoming flux. Indeed, if the irradiating flux is $L_{\rm w}^0 \vec{k}$ where $\vec{k}$ is a unit vector and the normal to the stellar surface is given by the unit vector $\vec{n}$ then the flux crossing the surface element is $L_{\rm w} = L_{\rm w}^0 \vec{k}\cdot\vec{n}$. The function $L_{\rm w}^0$ can take different forms depending of what the source of irradiation is assumed to be. It is common to assume symmetric direct heating from a point source, that is irradiation by a wind radially expanding from the pulsar, but it has been proposed that the wind might be reprocessed by an intra-binary shock \citep{romaniIBS} or channelled by the companion's magnetic field \citep{romaniBDuct} thus making $L_{\rm w}^0$ highly non-trivial in those cases.

There are several examples in the literature that show that the direct-heating model works well when fitting some optical light curves \citep[e.g.][]{van_kerkwijk_evidence_2011, breton_discovery_2013}. This tells us that i) the irradiating flux is, at least partly, deposited below the photosphere of the star as otherwise optical light curves would not be affected, and ii) that the deposition depth is probably shallow as otherwise heat would not emerge at the entry point on the photosphere. These two points have recently been backed in \citet{zilles_diagnosing_2019} who showed that only high-energy particles ($\gtrsim 100$MeV) can deposit their energy below the photosphere, and do so  at very shallow depths, typically after crossing a column density $ < 1000\si{g/cm^2}$.

\newcommand{\ir}{{\mathrm{ir}}}
%--------------------------------------------------------------------
\subsection{Basic transport equation}
In the following, we propose to supplement equation \eqref{eq:directheating} by adding the possibility of energy transport within a thin shell located just below the photosphere. The base of the shell is assumed to be unaltered by irradiation  which implies that it is much deeper than the reach of high-energy particles bombarding the star. The thickness of the shell must also be very small compared to the size of the star, and we will consequently consider it negligible.

Within this shell, we consider the stationary equation of conservation of energy,
\begin{equation}
\label{eq:consnrj}
	\vnabla \cdot \vj = \dot e
\end{equation}
where $\vj$ is the flux of energy per unit surface, and $\dot e$ is an external source of power per unit volume which, in the present case, is the irradiating power of the pulsar wind.  
Since this equation is linear in $\vj$, we may consider the base flux $\vb$ corresponding to the homogeneous solution, $\dot e = 0$, and a particular solution $\vi$ corresponding to irradiation such that
\begin{equation}
	\vj = \vb + \vi.
\end{equation}
The homogeneous solution $\vb$ is in principle part of a general solution of the full set of stellar-structure equations. Making the simplification that the star has a spherical photospheric surface of radius $R_*$, one has 
\begin{equation}
	b_r(R_*) =  \sigma_{\rm sb} T_{\rm b}^4,
\end{equation}
where $b_r$ is the radial component of $\vb$ and $T_{\rm b}$ is the base temperature, that is the photospheric temperature in absence of irradiation. This boundary condition is in fact all we need from the base solution for the following derivations.

In order to compute the particular solution $\vi$, we use the condition that the inner surface of the shell is unaffected by irradiation, which gives the boundary condition 
\begin{equation}
\label{eq:bcondir}
	i_r(R_{\rm i}) = 0,
\end{equation}
where $R_{\rm i}$ is the radius at the base of the shell. 

We now proceed to average equation \eqref{eq:consnrj} over the thickness of the shell. We start with integrating \eqref{eq:consnrj} over the volume of an element of shell corresponding to a surface $\delta S$ at the surface of the star between $R_*$ and $R_{\rm i}$. We immediately obtain $\int\dif{V} \dot e =  L_{\rm w} \delta S$ while we can apply Gauss' theorem to the divergence term on the left-hand side of Eq. \eqref{eq:consnrj} such that
\begin{equation}
\label{eq:div}
	\int \vi \cdot \dif{\vec{S}} = \int \dif{\vec{S}_{\parallel}}\cdot\vi_\parallel + \left(i_r(R_*) - i_r(R_{\rm i})\right)\delta S + \bigcirc\left(\frac{\Delta R}{R_*}\right), % \int_{R_{\rm i}}^{R_*} \dif{r}\oint \dif{\vec{C}}\cdot\vi_\parallel 
\end{equation}
where we have used the fact that $R_* - R_{\rm i} = \Delta R \ll R_*$ to approximate a shell element to a cylinder of height $\Delta R$ and cross-section $\delta S$, $\dif{\vec{S}_{\parallel}}$ is a surface element perpendicular to the lateral surface of this cylinder, and we have decomposed the energy flux into its radial and angular components $\vi = (i_r, \vi_\parallel)$.
% Therefore $\vi \cdot \dif{\vec{S}} = $ $\dif{\vec{C}}$ is a line element of the contour of that section, and we have decomposed the energy flux into its radial and angular components $\vi = (i_r, \vi_\parallel)$.

Defining the contour element $\dif{\vec{C}}$ such that $\dif{\vec{S}_{\parallel}} = \dif{r}\dif{\vec{C}}$ we can rewrite the parallel term of the right-hand side of equation \eqref{eq:div} as 
\begin{eqnarray}
\label{eq:paraterm1}
    \int \dif{\vec{S}_{\parallel}}\cdot\vi_\parallel & = & \int_{R_{\rm i}}^{R_*} \dif{r}\oint \dif{\vec{C}}\cdot\vi_\parallel, \\
\label{eq:paraterm2}
    & = & \int_{R_{\rm i}}^{R_*} \dif{r} \int \dif{S} \frac{1}{r} \vnablapara \cdot \vi_\parallel + \bigcirc\left(\frac{\Delta R}{R_*}\right),
\end{eqnarray}
where Eq. \eqref{eq:paraterm2} derives from Eq. \eqref{eq:paraterm1} by applying the two-dimensional Gauss' theorem within a section at radius $r$. Note that we have again neglected  the curvature of the surface as being of order $\bigcirc\left(\Delta R/R_*\right)$. Indeed, $r^{-1} \vnablapara$ is the angular part of divergence operator in spherical coordinates, 
% Now, we may revert Gauss' theorem in each 2-dimensional slice of the first term of the right-hand side of equation \eqref{eq:div},
% \begin{equation}
% \label{eq:slice}
% 	\oint \dif{\vec{C}}\cdot\vi_\parallel = \int \dif{S} \frac{1}{r} \vnablapara \cdot \vi_\parallel + \bigcirc\left(\frac{\Delta R}{R_*}\right),
% \end{equation}
% where $r^{-1} \vnablapara$ is the angular part of divergence operator in spherical coordinates, 
\begin{equation}
\label{eq:divpara}
\vnablapara = \frac{1}{\sin\theta}\left(\pderiv{\sin\theta}{\theta} \vuthet + \pderiv{}{\phi} \vuphi\right),
\end{equation}
where $(\theta,\phi)$ are respectively the colatitude and longitude at the surface of the star, and $(\vuthet,\vuphi)$ are the associated unit vectors.

Inserting equation \eqref{eq:paraterm2} back into equation \eqref{eq:div} using the boundary condition of equation \eqref{eq:bcondir}, and differentiating with respect to the surface elements $\delta S$ we obtain the averaged energy conservation equation,
\begin{equation}
\label{eq:consnrjav1}
	\vnablapara\cdot \int\dif{r} \frac{1}{r}\vi_\parallel = -i_r(R_*) +  L_{\rm w},
\end{equation}
where, in addition, we have used the fact that $\vnablapara$ is independent of $r$ to take it out of the integral on the left-hand side. 

Defining the ``average'' parallel energy flux as  
\begin{equation}
\label{eq:avparaflux}
\vJpara = \int_{R_{\rm i}}^{R_*} \dif{r} \frac{1}{r}\vi_\parallel(r),
\end{equation}
and introducing the irradiation temperature $\sigma_{\rm sb} T_\ir^4 = i_r(R^*)$, we may rewrite equation \eqref{eq:consnrjav1} as 
\begin{equation}
	\vnablapara\cdot \vJpara = -\sigma T_\ir^4 +  L_{\rm w}.
\end{equation}

The flux that escapes the star is given by $\vj_r(R_*) = \vb_r(R_*) + \vi_r(R_*)$. Since we have made the assumption that the irradiating power is thermalised before being re-radiated, this means that the escaped flux corresponds to a black-body at temperature $T_*$ such that 
\begin{equation}
	T_*^4 = T_{\rm b}^4 + T_\ir^4,
\end{equation}
and that this temperature corresponds to the actual temperature of the plasma at the photosphere.

This allows us to write our final superficial energy transport equation,
\begin{equation}
\label{eq:nrjredist}
\vnablapara\cdot \vJpara = - \left(\sigma_{\rm sb}\left(T_*^4 - T_{\rm b}^4\right) -  L_{\rm w}\right).
\end{equation}

One notes that if parallel energy transport can be neglected, that is $\vJpara = 0$, one naturally recovers the common direct heating of the companion star by the pulsar wind. In this case, $T_{\rm b}$ is directly equal to the night-side temperature of the star. Note that here we define the night-side temperature as the temperature at the point on the surface opposite to the pulsar's direction. This quantity is different from the effective temperature inferred at the inferior conjunction of the companion, which is an average over the visible surface at this particular phase.

\subsection{Transport laws: diffusion and convection}
We now consider that parallel energy transport follows a law of the form
\begin{equation}
\label{eq:transplaw}
	\vJpara = -\kappa \vnablapara T_* - T_* f(\theta)\sin\theta\vuphi,
\end{equation}
where the first term on the right-hand side accounts for diffusion-like effects and the second term for convection-like effects. The spherical coordinates are defined as for Eq. \eqref{eq:divpara} with the polar axis taken to be the spin axis of the star, and the prime meridian, $\phi = 0$, intersects the binary axis on the night side of the star. The parameter $\kappa$ is the diffusion coefficient with a dimension of energy per unit temperature per unit surface per unit time.

In the convection term, we consider that the surface temperature $T_*$ is convected by a velocity field that rotates around the angular-momentum axis of the star such that if the function $f$ is a constant then the convecting flow is in solid rotation with a velocity field $f\sin\theta\vuphi$. However, the polar convection profile $f(\theta)$ may be prescribed to reflect theoretical predictions such as, for instance, equatorial jets (see, e.g.,  \citet{showman_equatorial_2011} and below).

Here, we have assumed that the surface temperature $T_*$ is a good proxy for the transport properties of the shell. Indeed, if the energy is deposited at a shallow depth below the surface, then parallel temperature gradients should be maximum near the surface and so should be diffusion. Similarly, assuming a sufficiently smooth radial temperature profile in the shell then the photospheric temperature can be chosen as representative of convective transport. Nevertheless, the law of equation \eqref{eq:transplaw} should be considered as an effective description of the physics taking place in the outer shell of the star and not a as law derived from first principles.

To go further, we assume that $\kappa$ is a constant. Inserting equation \eqref{eq:transplaw} in the energy redistribution equation \eqref{eq:nrjredist}, we obtain 
\begin{equation}
\label{eq:nrjredist2}
	\kappa \vnablapara^2 T_* + f(\theta)\partial_\phi T_* = \sigma_{\rm sb}\left(T_*^4 - T_{\rm b}^4\right) - L_{\rm w},
\end{equation}
where $\vnablapara^2$ is the angular Laplacian.

Note that equation \eqref{eq:transplaw} is certainly not the only one possible solution but we favour it in this article owing to its relatively mathematical simplicity while retaining some of the expected qualitative behaviour. For instance, it could easily be generalised to more complex convection patterns and a non-constant $\kappa$.

We present in appendix \ref{ap:solution} a method to solve Eq. \eqref{eq:nrjredist2}. It is interesting to note that in many cases a good approximation can be obtained by linearising Eq. \eqref{eq:nrjredist2} around the direct heating solution of Eq. \eqref{eq:directheating} and decomposing $T_*$ onto spherical harmonics. We also found that this procedure can successfully be iterated in order to obtain the fully non-linear solution to Eq. \eqref{eq:nrjredist2}, thus providing a higher accuracy. We use the latter method in the rest of this article.

\section{Application }
\label{sec:application}
%pcp, effect of conv alone effect of diffusion alone, combined effect 
%combination with other reirradiation mechanisms, complementrarity 
% Non sphericity
% draghis 2019 for irradiation pattern centred on the equator
\subsection{Convection profiles \label{sec:convprof}}
The model of Eq. \eqref{eq:nrjredist2} depends on the choice of convection profile  $f(\theta)$ made by the modeller based on additional theoretical and/or empirical evidence. We have tried the following different forms,
\begin{eqnarray}
\label{eq:unifc}
   f(\theta ) & = & \nu, \\
  \label{eq:gaussfc}
    f(\theta ) & = & \nu \exp\left(-\frac{\theta^2}{2w^2}\right), \\
\label{eq:hermitefc}
    f(\theta ) & = & \exp\left(-\frac{\theta^2}{2w^2}\right) \sum_{i=0}^3 \nu_i  H_i\left(\frac{\theta}{w}\right),\\
\label{eq:bizonec}
   \quad f(\theta ) & = & +\nu \text{ if } |\theta| < w ; -\nu \text{ otherwise}.
\end{eqnarray}
In all these profiles, $\nu$ (or $\nu_i$) is the energy flux per unit temperature transported by convection.
Equation \eqref{eq:unifc} corresponds to a constant longitudinal advection, meaning that  if the properties of the superficial layer are constant across the entire surface (thickness, density, thermal capacity) then a constant $\nu$ corresponds to the constant angular velocity (around the spin axis of the star) of an advection flow in solid rotation around the star. If $\nu> 0$, then the flow is rotating in the same direction as the star on its orbit. Equation \eqref{eq:gaussfc} assumes that the flow is localised within a Gaussian belt of characteristic angular width $w$ around the equator. Equation \eqref{eq:hermitefc} is a generalisation of Eq. \eqref{eq:gaussfc} to an expansion into Hermite polynomials $H_n$ up to 3rd order. Indeed, such an expansion has been shown to be the eigen basis of the polar dependence of flow solutions to the shallow-water model developed in \citet{showman_equatorial_2011} for super-rotation in atmospheres of tidally-locked exoplanets. It follows that Eq. \eqref{eq:gaussfc} is simply Eq. \eqref{eq:hermitefc} with $\nu = \nu_0$ and $\nu_{i>0}=0$. Equation \eqref{eq:bizonec} corresponds to the particular case studied recently in \citet{kandel_atmospheric_2020} if diffusion is not included ($\kappa = 0$). In this model, a convection belt of width $2w$ is rotating around the equator while matter flows with opposite velocity at higher latitudes. 
All these profiles share the property that the convection pattern is dominated by an equatorial jet \citep[e.g.][]{showman_equatorial_2011}. We note that only Eqs. \eqref{eq:hermitefc} and \eqref{eq:bizonec} include the possibility of counter-rotating flows. 

\subsection{Heat redistribution maps}
\begin{figure*}
    \centering
    \includegraphics[width=\textwidth]{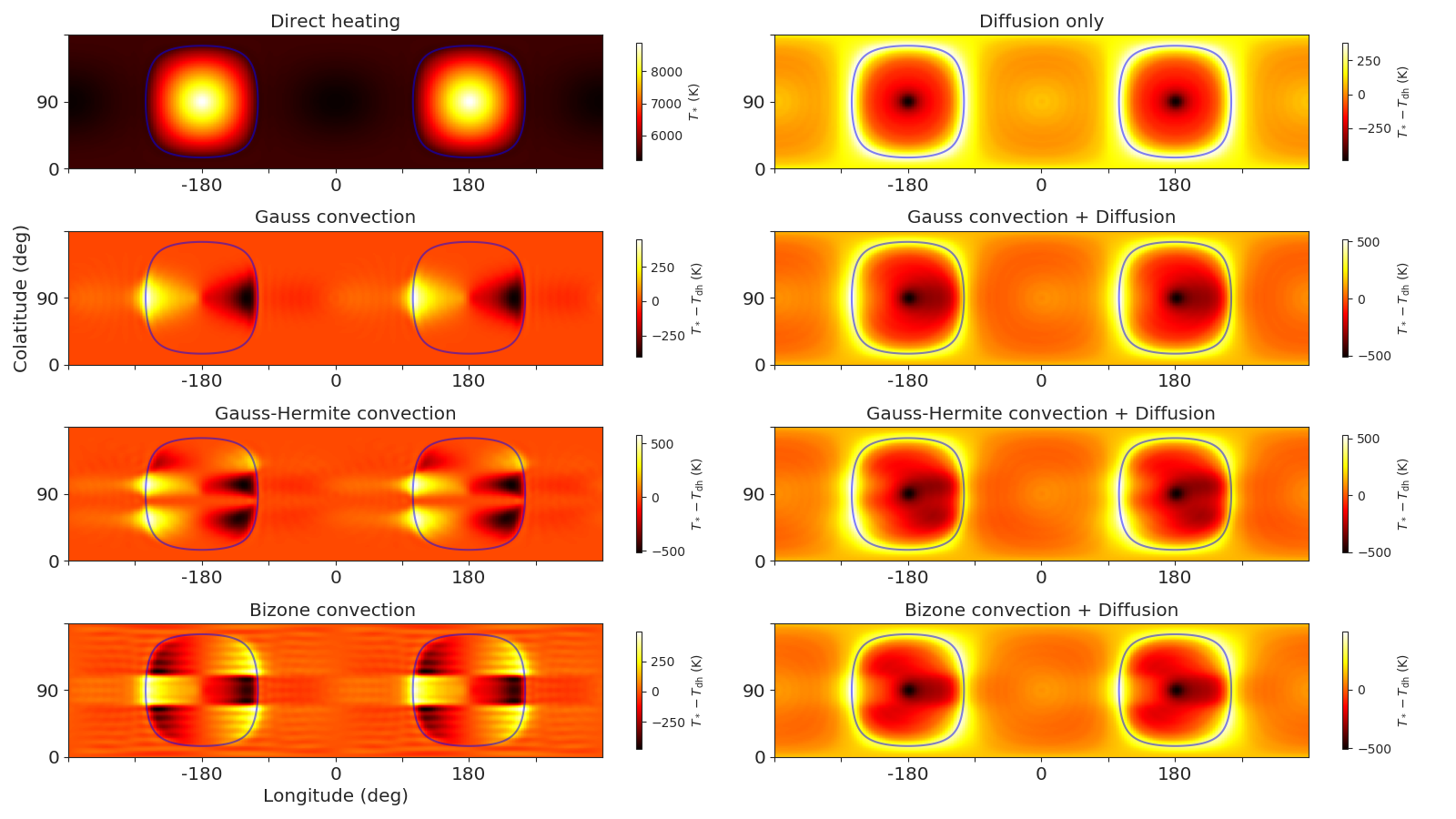}
    \caption{Temperature maps representing direct heating corresponding to the solution of Table \ref{tab:results} (top-left panel) and examples of temperature differences obtained when using the convection patterns of equations \eqref{eq:unifc}-\eqref{eq:bizonec} without (left column) or with (right column) diffusion. The convection and diffusion parameters are chosen to serve an illustrative purpose with $\kappa = 3500$, $\nu = 7000$, $\nu_i = \nu / 5^i$ and $w = 20^\circ$. The blue lines show the location of the terminator line of the corresponding direct heating. Two cycles of longitudes are shown for clarity. The point at longitude $180^{\circ}$, co-latitude $90^{\circ}$ faces the pulsar.}
    \label{fig:dt}
\end{figure*}

We show examples of the temperature difference with respect to direct heating, that is $T_* - \Tdh$, obtained using the above temperature profiles of Eqs. \eqref{eq:unifc}-\eqref{eq:bizonec} in Fig. \ref{fig:dt}. One sees that, in every case, the changes in temperature are located near the terminator of irradiation by the pulsar, as well as near the apex of the star in direction of the pulsar when diffusion is enabled. This is because it is where the strongest temperature gradients of the direct heating pattern are present. The additional wavy patterns that can be distinguished are due to the limited number of spherical harmonics used in the expansion of the solution. We have checked that for $l\geq 30$, these patterns entirely average out and do not bias the corresponding light curves (see next section).  

As can be seen in Fig. \ref{fig:dt}, diffusion transports energy from the day side to the night side symmetrically with respect to the binary axis (if the star is not spherical some small asymmetries can appear, in particular due to gravity darkening). 
On the contrary, the effect of convection is asymmetric between the leading and the trailing edge of the star, and localised at particular latitudes (except for the profile of Eq. \eqref{eq:unifc}). As a result, convection effectively creates hot and cold spots at the intersection of the characteristic latitude of a stream and the terminator line. However these spots are largely smoothed when diffusion is present. 

% In the present discussion, we will assume that convection is primarily driven by an equatorial jet around the star \citep{showman and others} and adopt the following polar convection profile which qualitatively embodies this idea,

% \begin{equation}
% 	f(\theta) = \nu\exp\left(-\frac{\theta^2}{2w^2}\right),
% \end{equation}
% where $\nu$ is the energy flux per unit temperature transported by convection, and $w$ is the characteristic width of the equatorial convection region. In the following we use $w=20\si{deg}$.

\subsection{Application to PSR J2215+5135}
As an example, the above model was fit to the multi-colour optical light curve (SDSS $g',r',i'$) of the redback companion of PSR J2215+5135 taken using the Auxiliary Port Camera (ACAM) mounted on the William Herschel Telescope in 2014. This data was initially presented in \citet{linares_peering_2018}, and is also employed in \citet{kandel_atmospheric_2020}. This is also the same object which was used by \citet{romaniIBS} to demonstrate the effectiveness in invoking an intra-binary shock in order to describe the asymmetries present in the light curve of PSR J2215+5135. 

The data were downloaded from the Isaac Newton Group of Telescopes Archive\footnote{\url{https://casu.ast.cam.ac.uk/casuadc/ingarch/}} along with associated flat fields and bias frames. The data were reduced using the ULTRACAM pipeline \citep{2007MNRAS.378..825D}. The instrumental magnitudes for PSR J2215+5135 were calculated via extraction of the source counts using an optimal photometry algorithm \citep{naylor98}. The counts for 6 surrounding non-variables stars which are in Data Release 1 from the Panoramic Survey Telescope and Rapid Response System (Pan-STARRS; \citealt{ps1}; \citealt{ps1db}) were also extracted and used to calibrate the magnitudes of PSR J2215+5135. The extracted light curve matches both the calibration and behaviour of that shown in \citet{linares_peering_2018}.

We have fitted the usual symmetric direct heating model without heat redistribution, and compared it with heat redistribution models using the convection profiles of Eqs.  \eqref{eq:unifc}-\eqref{eq:bizonec} both with and without diffusion, that is with $\kappa$ free or fixed to zero in Eq. \eqref{eq:nrjredist2}.
In each case, the parameter space was explored using \texttt{multinest}  (\citealt{MN1}; \citealt{MN2}; \citealt{MN3}) nested sampling algorithm as implemented in \textsc{python} through \texttt{pymultinest} \citep{pymultinest}. This algorithm was chosen because it allows to compare the evidence of each model, that is the probability of the model given the data, and therefore perform a direct comparison between them.

Apart from the heat redistribution parameters, we fit for extinction $E(g-r)$, the amplitude of the projected radial velocity $K_2$, distance $d$, base and irradiation temperatures $T_{\rm b}$ and $T_{\rm ir}$, Roche-lobe filling factor $f_{\rm RL}$ and system inclination $i$ \citep[e.g.][]{breton_koi_2012, breton_discovery_2013}. We also report in Table \ref{tab:results} some derived parameters of interest: the mass ratio $q$, the pulsar and companion masses $M_{\rm psr}$ and $M_{\rm c}$, and the irradiation efficiency $\epsilon$. The latter is defined as the ratio between the pulsar spin-down power and the irradiating power absorbed by the star \citep[e.g.][]{breton_discovery_2013}. To derive some of these parameters, we made use of the orbital characteristics obtained from pulsar timing, in particular the pulsar projected semi-major axis $a_p\sin i = 0.468141\pm0.000013\,$lt-s and the orbital period $P = 0.172502105\pm0.000000008\,$d \citep{2013ApJS..208...17A}.

\subsubsection{Priors \label{sec:priors}}
There are three parameters for which we set informed priors when exploring the parameter space with \texttt{Multinest}: the distance to the source, the optical extinction in the direction towards the source, and the radial velocity of the companion star. In addition, the inclination had a $\sin (i)$ prior applied, reflecting a uniform prior on the orientation of the system.

The distance prior has three components. The first is based on the estimated space density and transverse velocity of millisecond pulsars along the line of sight towards PSR J2215+5135, with the underlying spacial density for MSPs coming from \citet{2013MNRAS.434.1387L}. This component has a Gaussian distribution in distance from the Galactic centre with width $\sigma=4.5$ kpc, a decaying exponential in height above the Galactic plane with a scale height of 0.5 kpc, and a decaying exponential in transverse velocity, with a mean velocity of 100 km s$^{-1}$ \citep{2005AJ....129.1993M}. The second component comes from an upper limit on the system's parallax of $<1.8$ milliarcseconds at the 5$\sigma$ level, which was obtained from the second data release of the Gaia spacecraft \citep{GaiaDR2}. The third component comes from combining the most recent galactic electron density distribution model \citep{ymw17} with the dispersion measurement value of $69.1951\pm0.0002$ pc cm$^{-3}$ obtained from radio timing of PSR J2215+5135. The resulting prior is shown over the relevant parameter space in the distance plot of Fig. \ref{fig:mn}, and is not very constraining.

The prior on the optical extinction, that is $E({\rm g}-{\rm r})$, was a Gaussian centred on 0.13 and with width $\sigma=0.03$, and comes from the measured value from the \emph{Bayestar19} dust maps \citep{bayestar19}. The radial velocity of the companion star, $K_2$, had a Gaussian prior centred on 412 km s$^{-1}$ with width $\sigma=5$ km s$^{-1}$, inline with the estimated centre-of-mass velocity of the secondary given by \citet{linares_peering_2018}.

\subsubsection{Results}
\begin{figure*}
    \centering
    \includegraphics[width=\textwidth]{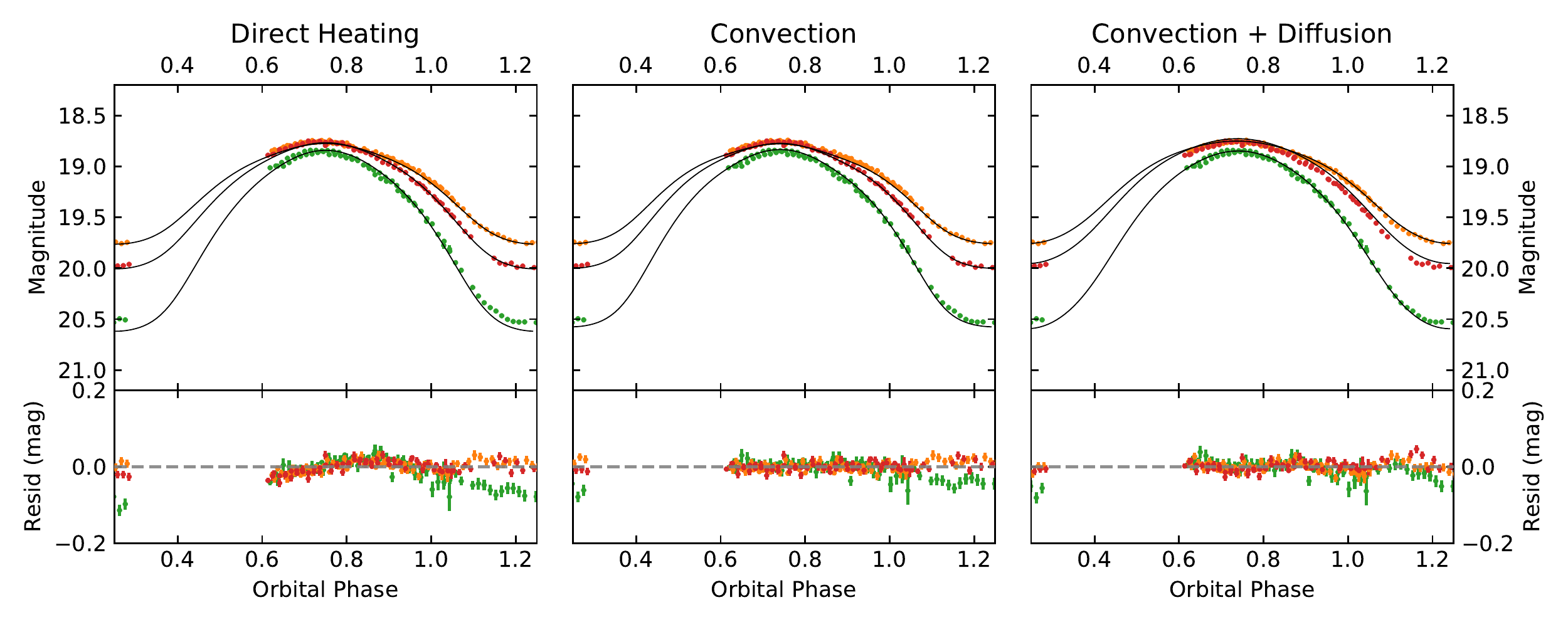}
\caption{The light curve of PSR J2215+5135 in SDSS \textit{g}$'$ (green), \textit{r}$'$ (red), and \textit{i}$'$ (orange) bands (top panel) and residuals after subtraction of the model light curve (bottom panel). In each case, the best posterior models is shown.}
    \label{fig:models}
\end{figure*}

It appears that models with the convection profiles of Eqs. \eqref{eq:gaussfc}-\eqref{eq:bizonec} all converge to the profile of Eq. \eqref{eq:unifc}. Indeed, their characteristic width is compatible with or larger than $w\sim\pi/2$. As a consequence, we report in detail only the results for the uniform convection model of Eq. \eqref{eq:unifc} with and without diffusion. We make an exception for the bizone convection profile of \eqref{eq:bizonec} without diffusion in order to compare with the recent work of \citet{kandel_atmospheric_2020}. We also report for comparison the symmetric direct heating model. The results of these four fits are collated in Table \ref{tab:results} by order of increasing evidence. The best-posterior light curves of these models are reported in Fig. \ref{fig:models}.

\begin{table*}
    \centering
    \renewcommand{\arraystretch}{1.2}
    \begin{tabular}{ccccc}
    \hline
        & Direct & Uniform & Bizone & Uniform + Diffusion \\
        \hline
        $\log Z$ & $-304$ & $-225.9$ & $-222.3$ & $-216.9$ \\
        $N_\mathrm{dof}$  & 229 & 228 & 227 & 227 \\
        $\chi^2_{\rm best\,\, likelihood}$ & $1101$ & $419.7$ &  $350.5$  & $464.2$ \\ 
        %$\chi^2$ best posterior & $1101$ & 412.4 &  $434.3$ med $489.3$ & $518.4$ med $584.4$ \\
        $\chi^2_{\rm median}$ & $1169$ & $481.2$ &  $371.4$ & $470.3$ \\
        %$376.9/231$ & $393.3/231$ \\
      %  $\ln \rm Posterior$ & $-624.5$ & $-468.7$ & $-467.2$ \\
      %  $\Delta \rm AIC$ & & &  \\
        \hline
        \multicolumn{5}{c}{Fitted parameters} \\
        \hline
        $E(g-r)$ & $0.122_{-0.048}^{+0.060}$ & $0.272_{-0.020}^{+0.011}$  & $0.288_{-0.018}^{+0.018}$ & $0.128_{-0.047}^{+0.040}$ \\
        $K_2$ (km/s) & $413_{-10}^{+9}$ & $413_{-7}^{+8}$ & $410_{-6}^{+10}$ & $412_{-10}^{+9}$ \\
        $d$ (kpc) &$3.33_{-0.25}^{+0.29}$ & $3.62_{-0.14}^{+0.11}$ & $3.63_{-0.15}^{+0.11}$ & $3.10_{-0.10}^{+0.11}$\\
        $T_{\rm b}$ (K) & $5596_{-157}^{+232}$ & $6582_{-184}^{+103}$ &$6814_{-245}^{+119}$ & $3451_{-2280}^{+1694}$ \\
        $T_\mathrm{ir}$ (K) & $7698_{-369}^{+566}$ & $9851_{-435}^{+223}$ & $10318_{-518}^{+291}$ & $9714_{-943}^{+479}$  \\
        $f_{\rm RL}$ & $0.861_{-0.018}^{+0.013}$ & $0.880_{-0.0086}^{+0.011}$ & $0.881_{-0.0012}^{+0.009}$ & $0.76_{-0.017}^{+0.027}$ \\
        $i \,\rm({}^\circ)$ & $64.4_{-8.1}^{+11.2}$ & $73.7_{-5.1}^{+8.4}$ & $82.3_{-9.9}^{+7.0}$ & $86.4_{-8.4}^{+3.5}$\\
        $\kappa$ (W/K/m${}^2$) & - & - & - & $53135_{-24128}^{+8400}$ \\
        $\nu$ (W/K/m${}^2$) & - & $4683_{-592}^{+604}$ & $5484_{-972}^{+669}$ & $9939_{-3627}^{+2335}$ \\
        $w$ (rad) & - & - & $2.1_{-1.1}^{+0.7}$ & - \\
        \hline
        \multicolumn{5}{c}{Derived parameters} \\
        \hline
        $q$ & $6.98_{-0.16}^{+0.16}$ & $6.98_{-0.12}^{+0.14}$ & $6.94_{-0.11}^{+0.17}$ & $6.96_{-0.16}^{+0.16}$\\
        $M_{\rm psr}$ ($M_\odot$) & $2.24_{-0.45}^{+0.64}$ & $1.86_{-0.17}^{+0.21}$ & $1.68_{-0.12}^{+0.24}$ & $1.65_{-0.11}^{+0.14}$ \\
        $M_{\rm c}$ ($M_\odot$) & $0.321_{-0.063}^{+0.090}$ & $0.267_{-0.024}^{+0.027}$ & $0.242_{-0.013}^{+0.032}$  & $0.237_{-0.010}^{+0.016}$ \\
        $\epsilon$ & $0.52_{-0.13}^{+0.23}$ & $1.23_{-0.21}^{+0.13}$ & $1.38_{-0.20}^{+0.19}$ & $1.08_{-0.36}^{+0.24}$ \\
        $T_{\rm N}^{\rm (spec)}$ (K) & $5462_{-149}^{+227}$ & $6396_{-177}^{+97}$ & $6614_{-231}^{+115}$ & $5728_{-138}^{+179}$\\
        $T_{\rm D}^{\rm (spec)}$ (K) & $7493_{-295}^{+477}$ & $9527_{-410}^{+219}$ & $10018_{-518}^{+291}$ & $7783_{-247}^{+336}$ \\
        \hline
    \end{tabular}
    \caption{Evidence sampling results for the three main models applied to J2215+5135: Direct heating, uniform convection without diffusion, and uniform convection with diffusion.
    $\log Z$ is the natural logarithm of the model evidence, and models are ranked by increasing evidence. $N_\mathrm{dof}$ is the number of degrees of freedom of each model, and we give the $\chi^2$ of the solution with the best likelihood (but not necessarily the best posterior probability) and of the median solution.  Model parameters are reported for the median solution with the 95\% confidence interval ($\pm 47.5\%$). \label{tab:results}}
\end{table*}

One sees that the most favoured model is the model with both uniform convection and diffusion, while uniform and bizone convection without diffusion yield quasi-identical solutions. 
Although the uniform and bizone convection models are compatible within error bars, one can see that their $\chi^2$ and evidence are sensibly different. We explain the latter by the additional parameter of the bizone model which allows for a better fit in part of the parameter space, and the former shows the sensitivity of the $\chi^2$ to the exact values of the parameters. Indeed, we could check that the uniform model yields the same $\chi^2=375.2$ as the bizone model when applied to the bizone median parameters reported in table \ref{tab:results}, as expected since $w>\pi/2$.

\begin{figure*}
    \centering
    \includegraphics[width=\textwidth]{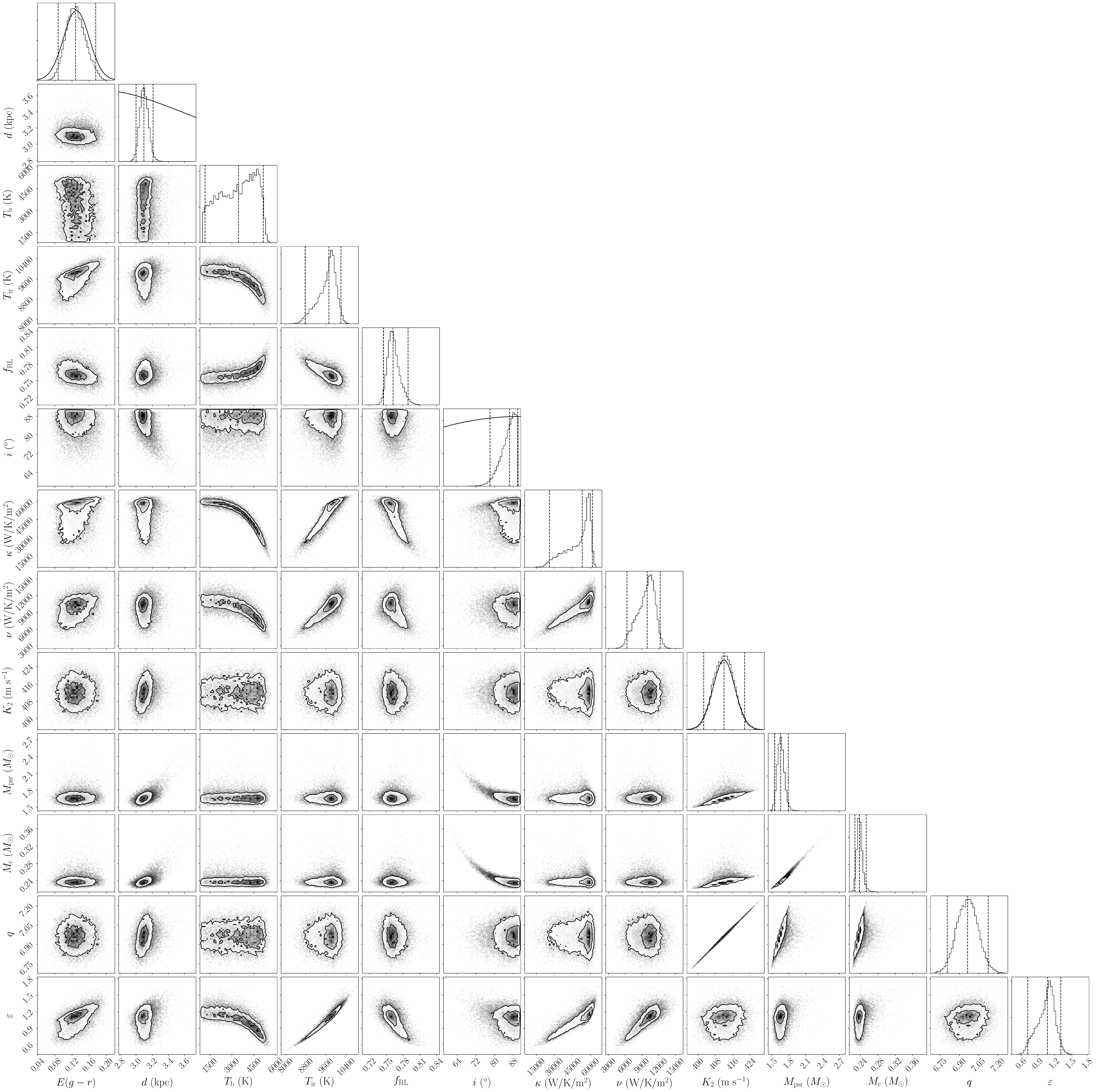}
    \caption{Results of the \textsc{multinest} fit of the light curve of PSR J2215+5135 using the uniform convection+diffusion model. The solid lines in the plots along the diagonal show the prior functions used, if a prior was specified.}
    \label{fig:mn}
\end{figure*}

According to their respective evidence, the uniform+diffusion model is respectively $\sim 8100$ times more likely than uniform convection alone and $\sim 220$ times more likely than bizone convection alone. However, one can see that the ranking in terms of best $\chi^2$ is quite different, reflecting the role of the priors in the results. As can be seen in Fig. \ref{fig:mn}, the uniform convection+diffusion model is the only one that fits the best within the distance and extinction priors, comparably to the direct heating model, while the two purely convective models stand at the edge of the extinction prior and require a significantly larger distance. 

\begin{figure}
    \centering
    \includegraphics[width= \columnwidth]{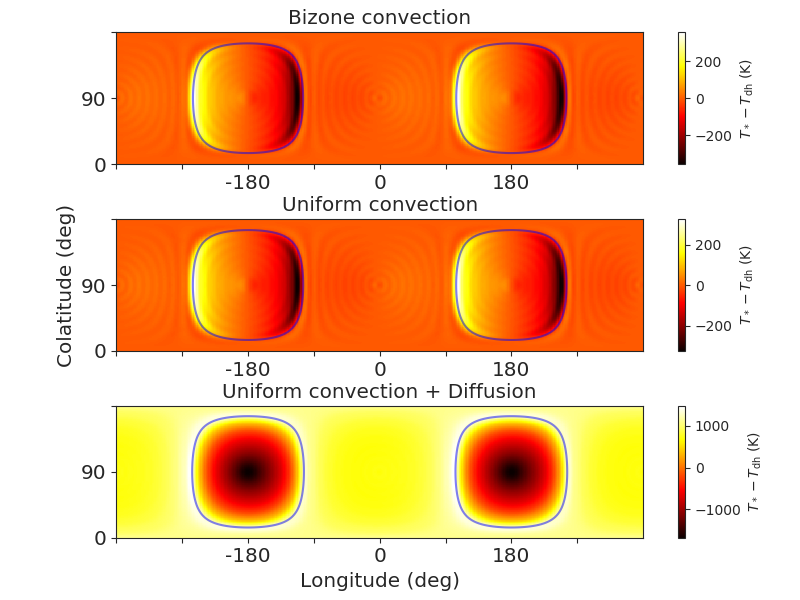}
    \caption{Maps of temperature difference with respect to direct heating for the best posterior parameters of the bizone convection model (top), the uniform convection model without diffusion (middle) and with diffusion (bottom) from the results of the light-curve fits of PSR J2215+5135's companion presented in Table \ref{tab:results}. The blue lines show the location of the terminator line of the corresponding direct heating pattern. Two cycles of longitudes are shown for clarity. The point at longitude $180^{\circ}$, colatitude $90^{\circ}$ faces the pulsar.}
    \label{fig:dtJ2215}
\end{figure}

\begin{figure}
    \centering
    \includegraphics[width= \columnwidth]{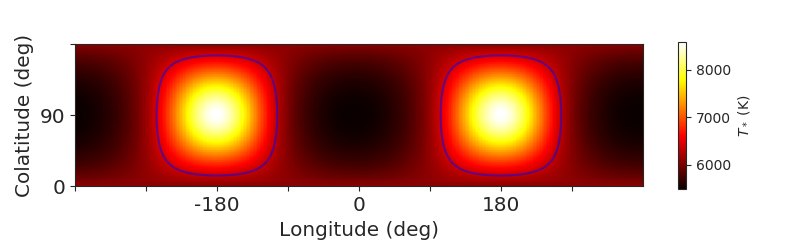}
    \caption{Temperature map of the best-posterior solution using with the uniform convection+diffusion model from the results of the light-curve fits of PSR J2215+5135 companion presented in Table \ref{tab:results}. The blue line shows the location of the terminator line of the corresponding direct heating pattern. Two cycles of longitudes are shown for clarity. The point at longitude $180^{\circ}$, colatitude $90^{\circ}$ faces the pulsar.}
    \label{fig:tempmapcd}
\end{figure}

This correlates with the fact that these models require both very high base temperature $T_{\rm b} \simeq 6550$K and irradiation temperature $T_{\rm ir} \simeq 9900$K implying a maximum day-side temperature over $T_{\rm D}^{(\max)} \sim (T_{\rm b}^4 + T_{\rm ir}^4)^{1/4}\simeq 11000$ K. Spectroscopic observations reported by \citet{linares_peering_2018} provide average night and day-side temperatures of $T_{\rm N}^{\rm (spec)} = 5660_{-380}^{+260}$ and $T_{\rm D}^{\rm (spec)} = 8080_{-280}^{+470}$K respectively. These temperatures are derived from spectra taken at inferior and superior conjunction of the companion respectively. These spectra result from the superposition of light originating from within the visible surface of the star which is not at a uniform temperature and therefore should be seen as average values. In particular, they are not equal to the minimum night-side temperature ($\sim T_{\rm b}$) and maximum day-side temperature $T_{\rm D}^{(\max)}$. We have estimated $T_{\rm N}^{\rm (spec)}$ and $T_{\rm D}^{\rm (spec)}$ for our models by computing the position of the peak of the spectrum resulting from the sum of the local black-body spectra of each visible surface elements at inferior and superior conjunction respectively.  The results, reported in Table \ref{tab:results} show that only the uniform convection+diffusion model, and with slightly more tension the symmetric direct heating model, are compatible with the spectroscopic observations of \citet{linares_peering_2018}, while the convection-only models require much larger temperatures for both sides of the star. We note the very important role of diffusion here. Indeed, diffusion simultaneously decreases the day-side temperature and increases the night-side temperature by $\sim 1000$ K compared to direct heating with the same parameters, as is shown on Fig. \ref{fig:dtJ2215}. It entails the much smoother temperature map of Fig. \ref{fig:tempmapcd} compared to, for instance, the direct heating model (top left panel of Fig. \ref{fig:models}) which allows a moderate day-night temperature difference despite the significantly larger irradiation temperature and cooler base temperature.

\section{Discussion}
\label{sec:disc}

\subsection{Base temperature}
Interestingly, the base temperature of the uniform convection+diffusion model is significantly lower than any of the other models with $T_{\rm b} = 3451_{-2280}^{+1694}$ K in comparison to at least $5500$K for the symmetric direct heating model and more than $6000$K for the convection-only models. Let us remember that the base temperature is the effective temperature that the star would have in absence of irradiation. Due to the lack of knowledge of the stellar structure of redback companions, it is difficult to know what this temperature should be in theory. The net result is that, especially for cases where the heat is substantially redistributed to the back of the star, the observed average night side temperature may depart significantly from the true temperature that it would have without irradiation.

It is however interesting to compare redback companions to companions of cataclysmic variables. Indeed, these stars have similar masses to redback companions, and similarly underwent Roche-lobe filling and mass transfer to the benefit of their primary (a white dwarf in this case). Nonetheless, these stars are not irradiated, letting their base temperature being seen, and have been largely studied both observationally and theoretically. One may therefore speculate that their effective temperature is similar to the base temperature of redback companions, in which case it would appear to be in the range $3000 - 4000$K depending on the mass of the star \citep[][]{knigge_evolution_2011}. The uniform convection+diffusion model is the only model compatible with this range.

\subsection{Interpretation of heat redistribution parameters}
The framework proposed in section \ref{sec:model} allows to redistribute energy at the surface of the star assuming a given transport law. However, determining such a law from first principle requires to determine not only the relevant microphysics but the hydrodynamical properties of stellar matter as well. This is out of the scope of the present paper and should be addressed in future work. Here we have focused on the study of the simplest possible convection-like and diffusion-like redistribution laws, by assuming only a latitudinal dependence for the convection profile (see Sect. \ref{sec:convprof}) and a constant diffusion coefficient. In the following we derive orders of magnitude to show that the values we obtain for the convection and diffusion parameters of the uniform convection+diffusion model applied to PSR J2215+5135's companion, $\nu$ and $\kappa$ respectively, can be compatible with some simple physical processes.

\subsubsection{Convection }
The fact that all convection profiles converged to a solution similar to the uniform rotation profile may simply mean that the finer details cannot be resolved with the available data. In particular, any latitudinal structure is bound to be largely if not completely averaged out since photometric information only provides the total flux contribution as a function of rotational phase with very little handle on the other axis. One may tentatively interpret the value of the convection parameter in terms of a wind velocity similar to the one-dimensional model of \citet{cowan_model_2011}. Thus, assuming uniform rotation of a shell of uniform column density $\Sigma$ at angular velocity $\omega$ one can write 
\begin{equation}
    \nu = \Sigma c_p \omega,
\end{equation}
where $c_p=  5k_b/3\mu \simeq 35000 \,\si{J.K^{-1}.kg^{-1}}$ is the specific heat capacity of a perfect gas at constant pressure, $k_b$ is Boltzmann's constant, and we have approximated the mean molecular mass $\mu$ to the mass of a proton. We may adopt the median value of the depth of maximum heat deposition, $500\si{g/cm^2}$ (see \citet{zilles_diagnosing_2019} and section \ref{sec:model}), as a fiducial value for $\Sigma$. This assumes that deeper layers are not affected by latitudinal convection. The characteristic hydrodynamical velocity is the speed of sound $c_s \sim \sqrt{k_b T_*/\mu}$ giving a fiducial $\omega = c_s/R_*$. It follows that 
\begin{equation}
    \nu = 5600\, \si{W.K^{-1}.m^{-2}} \left(\frac{\Sigma}{500\,\si{g/cm^2}}\right) \left(\frac{T_*}{8000\,\si{K}}\right)^{1/2}\left(\frac{0.37\,R_\odot}{R_*}\right),
\end{equation}
where we have derived the stellar radius $R_*$ from the result of the fit. Remarkably, the above fiducial value for $\nu$ agrees in order of magnitude with the results of Table \ref{tab:results}.

Interestingly, the bizone model does not reproduce the fit reported recently in \citet{kandel_atmospheric_2020} who uses the same model and the same photometric data. In particular, in \citet{kandel_atmospheric_2020} it is found that $w \simeq 35^{\circ}$ ($\theta_c$ in their notations) while we find $w > 90^{\circ}$, rendering the model virtually equivalent to uniform convection. It is unclear why this happens, however we note that the other major difference in the results of \citet{kandel_atmospheric_2020} is the relatively mild base temperature $T_{\rm b}$ ($T_N$ in their notations) as well as irradiation temperature. These quantities correlate with the level of extinction, and indeed one can see that the fitted values of $E(g-r)$ for our bizone and uniform models (Table \ref{tab:results}) are substantially into the tail of our prior on this parameter (see Sect. \ref{sec:priors}). On the other hand the fit of \citet{kandel_atmospheric_2020} assumes a fixed value of extinction (corresponding to the centre of our prior). Other reasons for the discrepancy might include their addition of a veiling flux (which we cannot assess with only photometric data) although they report that the inclusion of this extra component improves the fit without affecting the fitted parameters.

\subsubsection{Diffusion}
In the outer stellar envelope the main diffusion mechanism is radiative diffusion whose flux is $\vj_{\rm rad} = -(16/3)\sigma_{\rm sb} T^3 l_{\rm p} \vnabla T$ \citep[e.g.][]{kippenhahn_stellar_2012}, where $l_{\rm p}$ is the photon mean free path in the material and $\sigma_{\rm sb}$ is Stefan-Boltzmann's constant. Taking the average defined in equation \eqref{eq:avparaflux} of the radiative diffusion flux over a slab of stellar matter of height $H \ll R_*$, we can estimate 
\begin{equation}
\label{eq:jpararad}
    \vJpara{}_{\rm rad} \sim -\kappa_{\rm rad} \vnablapara T_*,
\end{equation}
where 
\begin{equation}
    \kappa_{\rm rad } \sim \frac{16}{3} \sigma_{\rm sb} T_*^3 l_{\rm p} \frac{H}{R_*^2}.
\end{equation}
The photon mean free path depends on the complex interplay of density, temperature and molecular composition \citep[e.g.][]{kippenhahn_stellar_2012}. Consequently, it is not possible to have a precise estimate of $l_p$ without a full modelling of at least the outer layer of the star. 

We note that $l_{\rm p} = (k\rho^{-1})$ where $\rho$ is the local density and $k$ the opacity of the material. In addition, $H \sim \Sigma /\rho$ where, as before, $\Sigma$ is the corresponding column density of the slab. Inserting a typical value for the opacity in Eq. \eqref{eq:jpararad}, that is $k \sim 1 \, \si{cm^2/g}$, we can estimate the density near the photosphere necessary to obtain a given value of the diffusion coefficient $\kappa_{\rm rad}$,
\begin{eqnarray}
        \rho & \sim & 2\times 10^{-8}\, \si{g/cm^3} \left(\frac{\kappa_{\rm rad}}{5\times10^4\,\si{W.K^{-1}.m^{-2}}}\right)^{-1/2} \\ & & \left(\frac{k}{1\,\si{cm^2/g}}\right)^{1/2}\left(\frac{\Sigma}{500\,\si{g/cm^2}}\right)^{1/2} \left(\frac{R_*}{0.37\,R_\odot}\right)^{-1}\left(\frac{T_*}{8000\,\si{K}}\right)^{3/2}. \nonumber
\end{eqnarray}
 %\sim \frac{2}{3}\frac{GM_*}{k R_*^2}\frac{\mu}{k_b T}
This value is an order of magnitude smaller than the photospheric density of Solar-type stars \citep[e.g.][]{vandenberg_use_2008} which have a similar surface gravity $\log g \simeq 4.4$ and temperature. However, this estimate of surface gravity does not take into account the effect of the star being close to filling its Roche lobe, which is bound to diminish the effective gravity near the surface of the star and in the atmosphere compared to the isolated case, thus diminishing the pressure and the density at the photosphere. We also note the higher temperature (compared to the Sun) on the day side of the companion, which would also tend to decrease the density at equal pressure. Thus, it seems possible that the value of the diffusion coefficient resulting from our fit can be explained with  radiative diffusion in the outer layers of the star, although a complete modelling of its atmospheric and sub-photospheric structure is necessary to answer this question with certainty.

\subsection{Orbital inclination and masses}
Pulsar timing of PSR J2215+5135 measures the projected semi-major axis of the pulsar as well as its orbital period. Using Kepler's third law, one combines these two parameters to compute the value of the so-called mass function \citep[e.g.][]{lyne_pulsar_2012} which relates the two masses of the system to the orbital inclination. In order to lift the degeneracy between masses and inclination, one needs two additional measurements. In the case of PSR J2215+5135, the mass ratio can be inferred from spectroscopic measurements of the companion's projected  radial velocity amplitude $K_2$, though with extra complications due to the irradiation effects \citep{linares_peering_2018}. On the other hand, fitting the optical light curve allows one to measure the inclination. 

Our results in Table \ref{tab:results} show that this quantity is highly model-dependent, ranging from ${64.4_{-8.1}^{+11.4}}^\circ$ for the direct heating model to ${86.4_{-8.4}^{+3.5}}^\circ$ for the diffusion+uniform convection model through somewhat intermediate values for the two convection-only models. Accordingly, the pulsar mass ranges from $2.24_{-0.45}^{+0.64} \rm{M_\odot}$ to $1.65_{-0.11}^{+0.14} \rm{M_\odot}$ for the direct heating and diffusion+uniform convection respectively. The direct heating values confirm those found in \citet{linares_peering_2018} using the same dataset, while being hardly compatible at the 95\% level with the diffusion+uniform convection values. Interestingly, the latter gives a similarly high inclination to what was found in \citet{romani_keck_2015} (see \citet{linares_peering_2018} for a review of previous measurements). In \citet{kandel_atmospheric_2020}, it is however argued that this previous result might have been biased by an extra blue veiling flux at the epoch of the observations of \citet{schroeder_observations_2014} (whose optical light curve they use), as suggested by a corresponding excessive night-side temperature of that fit compared to the spectroscopic constraints of \citet{linares_peering_2018}. 

In the present work, we see that inclination is substantially changing from one model to another, everything else being equal. Although we cannot here assess with certainty the presence of a veiling flux for lack of spectroscopic observations, the night-side temperature of the diffusion+uniform convection fit is not excessively large as discussed in Sect. \ref{sec:application}. Another possible caveat is the lack of a significant portion of the orbital light curve (see Fig. \ref{fig:models}), which might bias the fit especially considering the asymmetry of the light curve. We therefore conclude that a thorough investigation involving simultaneous spectroscopy and photometry across an entire orbit is desirable in order to be able to reduce the risk of bias.

\section{Conclusions}

In this paper we have considered the effects of heat redistribution at the surface of companion stars of spider pulsars. In effect, we have supplemented the usual direct heating model of irradiation, Eq. \eqref{eq:directheating}, with a single extra term accounting for the divergence of the heat flux within the stellar surface, Eq. \eqref{eq:nrjredist}. This may be seen as the simplest addition possible to direct heating models. On the other hand, the heat flux itself requires a complex modelling of the outer layer and atmosphere of the star combining microphysics, hydrodynamics and thermodynamics which is outside of the scope of the present work. 

In the spirit of studying the simplest possible extensions to direct heating models we have evaluated the effect of simple convection-like and diffusion-like laws, Eqs. \eqref{eq:transplaw} and \eqref{eq:nrjredist2}. The solution of the redistribution equation can be represented under the form of heat redistribution temperature maps, Fig. \ref{fig:dt}, which show that both convection and diffusion effects are most intense near the irradiation terminator or near the apex of the star towards the pulsar. Interestingly, convection is naturally able to produce patterns akin to hot or cold spots at the terminator. We also note that heat redistribution models are compatible with other models which modify the irradiation pattern such as intra-binary shock models \citep{romaniIBS} or magnetic-field ducting models \citep{romaniBDuct}, and with models that modify the base temperature such as hot and cold spots \citep[e.g.][]{shahbaz_properties_2017}.

We have applied our models to the light curve of the already well-studied companion of the redback pulsar PSR J2215+5135 in order to determine empirically the most likely form of the heat flux. Various convective flows with and without diffusion were tried. We found that, although every redistribution model provides a substantially better fit than the symmetric direct heating model, the model associating diffusion to convective flows in uniform rotation is most likely (see Table \ref{tab:results}). 

However, since with every model substantial fit residuals remain these results should be taken with caution and we consider that the main value of the different fits lies in the comparison with each other. Indeed, as it appears in Table \ref{tab:results}, the various models lead to sometimes very discrepant fitted parameters, in particular concerning the base and irradiation temperatures, the inclination, the filling factor or the irradiation efficiency. This suggests that, on top of detailed modelling, the determination of the ``true'' model of heat redistribution will certainly require complementary observations such as spectroscopic measurements of the effective temperature, or accurate and independent distance measurements.

\section*{Acknowledgements}

The authors acknowledge support of the European Research Council, under the European Union's Horizon 2020 research and innovation program (grant agreement No. 715051; Spiders). This research made use of Astropy,\footnote{\url{http://www.astropy.org}} a community-developed core Python package for Astronomy \citep{astropy:2013, astropy:2018}.

\section*{Data availability}
The raw data on PSR J2215+5135 were obtained from the Isaac Newton Group of Telescopes Archive\footnote{\url{https://casu.ast.cam.ac.uk/casuadc/ingarch/}}. The reduced light curves used in this paper were produced using the ULTRACAM pipeline \citep{2007MNRAS.378..825D}, and are available from a permanent public repository\footnote{\url{https://doi.org/10.5281/zenodo.3894748}}.

%
%The Acknowledgements section is not numbered. Here you can thank helpful
%colleagues, acknowledge funding agencies, telescopes and facilities used etc.

%%%%%%%%%%%%%%%%%%%%%%%%%%%%%%%%%%%%%%%%%%%%%%%%%%



\bibliographystyle{mnras}
\bibliography{Heat_redistribution} % if your bibtex file is called example.bib

%\begin{thebibliography}{99}
%\end{thebibliography}

%%%%%%%%%%%%%%%%%%%%%%%%%%%%%%%%%%%%%%%%%%%%%%%%%%

%%%%%%%%%%%%%%%%% APPENDICES %%%%%%%%%%%%%%%%%%%%%
%
\appendix
\section{Solution of the heat redistribution equation }
\label{ap:solution}
In this section we use the lighter notation $T\equiv T_*$.
\subsection{Solution of the linearised transport equation}\label{sec:linsol}
Equation \eqref{eq:nrjredist} is strongly non-linear in $T$ due to the $T^4$ term. However, given the relative success of direct heating models we may assume that energy redistribution is only a perturbation of the temperature distribution at the surface of the star. Thus, we may write $T = \Tdh + t$ and, assuming $t \ll \Tdh$, expand equation \eqref{eq:nrjredist} to first order in $t$, 
\begin{equation}
\label{eq:nrjredistlin}
  a t - \kappa \vnablapara^2 t - f(\theta)\partial_\phi t =  s,
\end{equation}
where 
\begin{eqnarray}
	a & = & 4 \sigma_{\rm sb} \Tdh^3 , \\
	s & = & \kappa \vnablapara^2 \Tdh + f(\theta) \partial_\phi \Tdh.
\end{eqnarray}

The linearised equation \eqref{eq:nrjredistlin} can be solved algebraically after decomposing the function onto the orthogonal basis of spherical harmonics $\{Y_{lm}\}_{l \geq 0 ; -l \leq m \leq l}$ \citep[e.g.][]{olver_nist_2010}. In this basis, the functions $t, a$ and $s$ are represented by the vectors $\vec{t}, \vec{a}$ and $\vec{s}$ respectively, such that 
\begin{equation}
	t = \sum t_{lm} Y_{lm},
\end{equation} 
where $t_{lm}$ are the coefficients of $\vec{t}$, and similarly for $s$, $a$ and $f$. It follows that equation \eqref{eq:nrjredistlin} can be expanded into
\begin{eqnarray}
\label{eq:nrjredistylm}
	\sum_{lml'm'} t_{lm} a_{l'm'} Y_{lm} Y_{l'm'} + \kappa \sum_{lm} l(l+1) t_{lm} Y_{lm} & & \nonumber \\
	- \sum_{lml'm'} i m t_{tm} f_{l'm'} Y_{lm}Y_{l'm'} & = & \kappa \sum_{lm} s_{lm} Y_{lm}.
\end{eqnarray}   
By projecting equation \eqref{eq:nrjredistylm} onto each spherical harmonic we obtain a set of linear algebraic equations the solution of which is formally given by
%\begin{equation} 
%	\sum_{lml'm'} t_{lm} \mu_{LMlml'm'} a_{l'm'}  + \kappa L(L+1) t_{LM} Y_{LM} = \kappa s_{LM}.
%\end{equation}
\begin{equation}
\label{eq:solution1}
	\vec{t} =  M^{-1} \vec{s},
\end{equation}
where we have introduced the matrix $M= \{M_{ij}\}$. Its coefficients are defined by 
\begin{equation}
\label{eq:solution2}
	M_{\alpha\beta} = \sum_{\gamma}  \mu_{\alpha\beta\gamma} (a_{\gamma} - i m_\beta f_\gamma) + \kappa l_\alpha(l_\alpha+1)\delta_{\alpha\beta}, %+ m \delta_{ij},
\end{equation}
where $\delta_{\alpha\beta} = 1$ if $\alpha=\beta$ and $0$ otherwise, and each index $\alpha,\beta,\gamma$ maps onto a different pair of spherical-harmonic indices  $(l,m)$ (for example $\alpha=\{0,1,2,3...\} \rightarrow (l_\alpha,m_\alpha) = \{(0,0), (1,-1), (1,0), (1,1)...\}$). We have introduced the spherical-harmonic multiplication coefficients $\{\mu_{\alpha\beta\gamma}\}$ such that 
\begin{equation}
	Y_{\beta}Y_{\gamma} = \sum_\alpha  Y_{\alpha}\mu_{\alpha\beta\gamma}.
\end{equation}
These coefficients can, for example, be obtained from the Clebsch-Gordan coefficients \citep[e.g.][]{olver_nist_2010}. Alternatively, one can compute them numerically using publicly available tools such as \texttt{shtools} \citep{wieczorek_shtools_2018} \footnote{\url{https://shtools.github.io/SHTOOLS/}}.

\subsection{Solution of the full non-linear transport equation}
Some irradiated stars show very large temperature differences between their day and night sides, to the point that the temperature difference might exceed the temperature of the night side. In this case, the assumption that heat redistribution is only a perturbation of direct heating may fail. Here, we propose a fixed-point scheme to solve the full non-linear equation \eqref{eq:nrjredist2} by iterating the linearised solution of section \ref{sec:linsol}.

At each iteration, the temperature distribution $T_{n+1}$ is calculated according to
\begin{equation}
\label{eq:Tn}
	T_{n+1} = T_n + t_{n+1},
\end{equation}
where $t_{n+1}$ is the solution of equation \eqref{eq:nrjredist2} linearised with respect to $T_n$,
\begin{equation}
\label{eq:tn}
	A_n t_{n+1} - \kappa \vnablapara^2t_{n+1} - f(\theta) \partial_\phi t_{n+1} = S_{n},
\end{equation}
where 
\begin{eqnarray}
	A_n & = & 4 \sigma T_n^3 , \\   
	S_n & = & \kappa \vnablapara^2 T_n + f(\theta) \partial_\phi T_n - \sigma(T_n^4 - T_{\rm b}^4) + L_{\rm w}.
\end{eqnarray}

Equations \eqref{eq:Tn} and \eqref{eq:tn} form a sequence that can be initialised with $T_0 = \Tdh, t_0 = 0$ such that $t_1$ is equal to $t$ of the previous section. The solution of equation \eqref{eq:tn} is given by equations \eqref{eq:solution1} and \eqref{eq:solution2} only replacing the vectors $\vec{a}, \vec{s}, \vec{t}$ by the corresponding $\vec{a}_{n}, \vec{s}_{n}, \vec{t}_{n+1}$.

In practice, this scheme converges after a few iterations with the stopping criterion $\|\vec{t}_{n+1}\| < 1$K.

%%%%%%%%%%%%%%%%%%%%%%%%%%%%%%%%%%%%%%%%%%%%%%%%%%

% Don't change these lines
\bsp	% typesetting comment
\label{lastpage}
\end{document}